\documentclass[letter]{aa}  

\usepackage{graphicx}
\usepackage{amsmath}
\usepackage{txfonts}

\newcommand{\kms}{\text{km~s}^{-1}}
\newcommand{\kpc}{\text{kpc}}
\newcommand{\ve}{v_\text{e}}
\newcommand{\vetr}{v_\text{e,340}}
\newcommand{\vei}{v_{\text{e},\infty}}
\newcommand{\se}{\sigma_\text{e}}
\newcommand{\vcut}{v_\text{cut}}
\newcommand{\de}{\text{d}}

\begin{document}

\title{The escape speed curve of the Galaxy obtained from Gaia DR2 implies a heavy Milky Way}

\author{
G. Monari \inst{1} \and 
B. Famaey \inst{2} \and 
I. Carrillo \inst{1} \and
T. Piffl \inst{1} \and
M. Steinmetz \inst{1} \and 
R. F. G. Wyse \inst{3,4} \and
F. Anders \inst{1} \and 
C. Chiappini \inst{1} \and
K. Jan{\ss}en \inst{1}
}

\institute{
Leibniz-Institut fuer Astrophysik Potsdam (AIP)
An der Sternwarte 16, 14482 Potsdam, Germany\\  
\email{gmonari@aip.de}
\and
Universit\'{e} de Strasbourg, Observatoire astronomique de Strasbourg, CNRS UMR 7550, 11 rue de l'Universit\'{e}, 67000 Strasbourg, France
\and
Department of Physics \& Astronomy, Johns Hopkins University, Baltimore, MD 21218, USA
\and 
Leverhulme Trust Visiting Professor, University of Edinburgh, UK
}

\date{Received xxx; accepted yyy}

\abstract
{We measure the escape speed curve of the Milky Way based on the analysis of the velocity distribution of  $\sim 2850$ counter-rotating halo stars from the Gaia Data Release 2. The distances were estimated through the \texttt{StarHorse} code, and only stars with distance errors smaller than 10\% were used in the study. The escape speed curve is measured at Galactocentric radii ranging from $\sim 5$~kpc to $\sim 10.5$~kpc. The local Galactic escape at the Sun's position is estimated to be $\ve(r_\odot)=580 \pm 63~\kms$, and it rises towards the Galactic centre. Defined as the minimum speed required to reach three virial radii, our estimate of the escape speed as a function of radius implies for a Navarro-Frenk-White profile and local circular velocity of $240 \, \kms$ a dark matter mass $M_{200}=1.28^{+0.68}_{-0.50} \times 10^{12} M_\odot$ and a high concentration $c_{200}=11.09^{+2.94}_{-1.79}$. Assuming the mass-concentration relation of $\Lambda$CDM, we obtain $M_{200}=1.55_{-0.51}^{+0.64}\times 10^{12}~M_\odot$ and $c_{200}=7.93_{-0.27}^{+0.33}$ for a local circular velocity of $228 \, \kms$.}


\maketitle

\section{Introduction}

While much progress has been made in recent years regarding our understanding of the formation of galaxies, the very basic fundamental parameters of our own Galaxy are still very poorly known. Of these fundamental parameters, the mass of our Galaxy is perhaps the most important, and it is still unknown within a factor of four \citep{BG2016}. This parameter is of particular interest because it provides a test for the current $\Lambda$ cold dark matter ($\Lambda$CDM) paradigm, and the validity of recipes such as abundance matching \citep[e.g.][]{Behroozi}.

The primary objective of the cornerstone Gaia mission \citep{GaiaDR2} is to allow us to build a detailed dynamical model of the Galaxy, including all of its components, and to provide us insight into its structure, its formation, and its evolutionary history. Before building such a detailed and exhaustive model, we can already take advantage of the accuracy, quality, and extent of the data provided by the Gaia Data Release 2 (DR2) to determine the fundamental parameters of the Galaxy such as its virial mass more accurately than ever before. Several recent studies have tried to estimate this parameter through various methods: \citet{Fritz2018} found that a virial mass of at least $1.6 \times 10^{12} M_\odot$ would be needed to keep a reasonable number of satellite dwarf galaxies bound. \citet{Hattori2018} assumed that extremely high velocity stars with chemical properties similar to those of the Galactic halo are bound to the Galaxy and showed that the virial mass should be higher than $1.4 \times 10^{12} M_\odot$. \cite{HawinsWyse2018} also studied a sample of hyper-velocity stars and found that their chemistry is compatible with that of the stellar halo. \citet{Watkins2018} used a mass estimator based on the kinematics of halo globular clusters to find a virial mass of $1.67^{+0.79}_{-0.50} \times 10^{12} M_\odot$. Finally, \citet{Posti2018} also used the kinematics of globular clusters to fit a two-component distribution function in action space as well as the Galactic potential including a prolate Navarro-Frenk-White (NFW) halo, to find a slightly lower value for the virial mass.

Here, our new estimate of the Milky Way mass will be made through measuring the escape speed curve of the Galaxy at Galactocentric radii ranging from $\sim 5$~kpc to $\sim 10.5$~kpc based on Gaia DR2. Since the escape speed directly measures the difference between the local potential and that in the outskirts of the Galaxy, it allows us to constrain its total mass based on local measurements. In doing this, we follow up on the previous studies by \citet{LeonardTremaine1990}, \citet{Kochanek96}, \citet{Smith2007}, and \citet[][hereafter P14]{Piffl2014}.

The general idea is to use a simple model for the tail of the velocity distribution of halo stars at a given radius, based on a truncated power law \citep{LeonardTremaine1990} motivated by simulations \cite[][P14]{Smith2007}. Based on a model for the baryonic distribution of the Milky Way, the virial mass of the Galaxy can be adjusted to best fit the escape speed curve. In Section~\ref{sec:data} we present our selection of Gaia DR2 stars. The method for estimating the escape speed is then presented in Section~\ref{sec:Methods}, and the results are summarized in Section~\ref{sec:results1}. The corresponding virial mass of the Galaxy dark matter halo is presented in Section~\ref{sec:results2}, and we conclude in Section~\ref{sec:conc}.

\section{Data}\label{sec:data}
We used stars from the Gaia DR2 with line-of-sight (l.o.s.) velocity information. Following  \cite{Marchetti2018}, we selected only stars with the following Gaia flags: 
\begin{itemize}
\item $\texttt{ASTROMETRIC\_GOF\_AL}<3$;
\item $\texttt{ASTROMETRIC\_EXCESS\_NOISE\_SIG}\leq 2$;
\item $-0.23 \leq \texttt{MEAN\_VARPI\_FACTOR\_AL} \leq 0.32$;
\item $\texttt{VISIILITY\_PERIODS\_USED}>8$;
\item $\texttt{RV\_NB\_TRANSITS} > 5$.
\end{itemize}
We used new distance estimates (Anders et al. in prep.) obtained with the \texttt{StarHorse} code \citep{Queiroz2018}, which combines multiband photometric information (APASS, 2MASS, and AllWISE) and the Gaia astrometric information with a Bayesian approach, accounting for the global Gaia DR2 parallax zero-point shift of -0.029~mas \citep{Lindegren2018,Arenou2018}. Based on these distances and on the Gaia DR2 astrometry and l.o.s. velocities, we computed the Galactocentric cylindrical positions $(R,z,\phi)$ and velocities $(v_R,v_\phi,v_z)$ of these stars. For this, we assumed a distance of the Sun from the centre of the Galaxy of $r_\odot=8.34~\kpc$, a circular velocity at the Sun of $v_\text{c}(r_\odot)=240~\kms$ \citep{Reid2014}, and a peculiar solar motion of $(U_\odot,V_\odot,W_\odot)=(11.1,12.24,7.25)~\kms$ , as estimated by \citet{Schonrich2010}. Of these stars we selected only those that counter-rotate, that is, stars with $v_\phi<0$, in order to ensure that the kinematic tail is representative of stellar halo-type stars. We checked that reflecting $v_\phi$ for these stars around $v_\phi=0$, we obtain the kind of kinematics typical of the stellar halo, with velocity dispersions of $\sim 140~\kms$ \citep[e.g.][]{Battaglia2005}.

Of these stars, we selected only those with distance errors smaller than 10 per cent. We then computed their Galactocentric spherical distance $r=\sqrt{R^2+z^2}$ and the speed $v=\sqrt{v_R^2+v_\phi^2+v_z^2}$. However, we did not simply combine the observable quantities, but used a Monte Carlo technique (explained below), which allowed us to achieve better estimates of the $r$ and $v$ uncertainties than a simple propagation of errors.

\section{Methods}\label{sec:Methods}
As in P14, we assumed that the velocities $v$ of stars in the kinematic tail of the halo are distributed according to a power law, with a truncation at the escape velocity $\ve$ \citep{LeonardTremaine1990},
\begin{equation}\label{eq:f}
f(v|\ve,k)=\begin{cases}
    (k+1) (\ve-v)^k / (\ve-\vcut), & v\leq\ve,\\
    0, & v>\ve,
  \end{cases}
\end{equation}
where $\vcut$ corresponds to a lower cut in $v$ that defines the fast stars that should be distributed as in equation~\eqref{eq:f}. We note that $\ve$ and $k$ depend in general on the position in the Galaxy. Both $\vcut$ and typical values of $k$ can be read off simulations. In particular, P14 have shown from simulations that the typical range for $k$ is $2.3<k<3.7$ and that stars follow this power law for $\vcut \sim 250~\kms$.

The Bayes theorem can be used to show that the probability of the model parameters $(\ve,k)$, given $N$ stars with velocity determination $v_i$ (for the $i$-th star) is 
\begin{equation}\label{eq:P}
P(\ve,k|v_{i=1,...,N})=\frac{P(\ve)P(k)\Pi_{i=1}^{N} f(v_i|\ve,k)}{\int\int P(\ve)P(k)\Pi_{i=1}^{N} f(v_i|\ve,k)~\de\ve~\de k},
\end{equation}
where the uncertainty on the determination of $v_i$ is neglected for the moment, and the denominator is the same for all the points in the $(\ve,k)$ space, so that we do not need to determine it. For the {\it \textup{a priori}} probabilities we chose $P(\ve)\propto 1/\ve$ and $P(k)$ uniform, as in \cite{LeonardTremaine1990}.

Since we are interested mostly in $\ve$, we finally marginalized along $k$ to obtain the probability density of $\ve$,
\begin{equation}\label{eq:vemarg}
P(\ve|v_{i=1,...,N})=\int_{2.3}^{3.7} P(\ve,k|v_{i=1,...,N})~\de k,
\end{equation}
where the range of marginalisation is obtained from P14, as explained above.

However, at this stage, equation~\eqref{eq:P} does not consider the uncertainty on the individual stellar parameters ($r$ and $v$). 
Hence, in order to obtain a robust result, we used a Monte Carlo or bootstrap technique, repeating the scheme detailed below 100 times.
\begin{itemize}
\item For each star in the sample, we drew a random distance, proper motion, and l.o.s. velocity from Gaussians with means and dispersions (the quantity and its uncertainty) given by the \texttt{StarHorse} code and the Gaia catalogue; from these, we computed $r$ and $v$ assuming for the transformation the same Galactic parameters as were used in Section~\ref{sec:data}, and negligible uncertainties on the positions on the sky of the stars.
\item We used only stars with drawn random distance smaller than $5~\kpc$ in order to have a sample of stars with Bayesian distance estimates dominated by the Gaia parallax information, and not by the photometry;  considering the parallax shift, 99\%\ of the stars in Gaia DR2 with l.o.s. velocities and parallax errors better than 10\%\ are found inside $\sim 5~\kpc$, and inside the same sphere, 90\%\ of the stars have parallax error smaller than $\sim 10$ \%; the typical size of the samples obtained in this way is $\sim 2850$ stars.
\item We resampled the data set obtained in this way with replacement, so that each time some stars are randomly excluded and some stars are repeated more times in the sample. 
\item We binned the data into bins of size $0.6~\kpc$, with the innermost bin centred at  $r=5.34~\kpc$ and the outermost at $r=10.14~\kpc$.
\item For the $i$-th iteration (and resampling), we calculated for each bin the escape velocity $\ve^i$ and its uncertainty $\se^i=\max(\ve^i-v_{16}^i,v_{84}^i-\ve^i)$, where $v_{16}^i$ and $v_{84}^i$ are the 16th and 84th percentiles of the $\ve$ distribution.
\end{itemize}
We then estimated $\ve$ for each $r$-bin as the median of the $\ve^i$ estimates of each iteration, and its uncertainty as the sum in quadrature of the square root of the mean $\se^{i \, 2}$ and of the standard deviation of the values $\ve^i$. The error bars obtained in this way contain a systematic component because the top of the error bar for each bin always depends critically on the fastest star that can be included in the sample in one of the 100 Monte Carlo realisations.

\section{Escape speed curve}\label{sec:results1}

\begin{figure*}
\includegraphics[width=\textwidth]{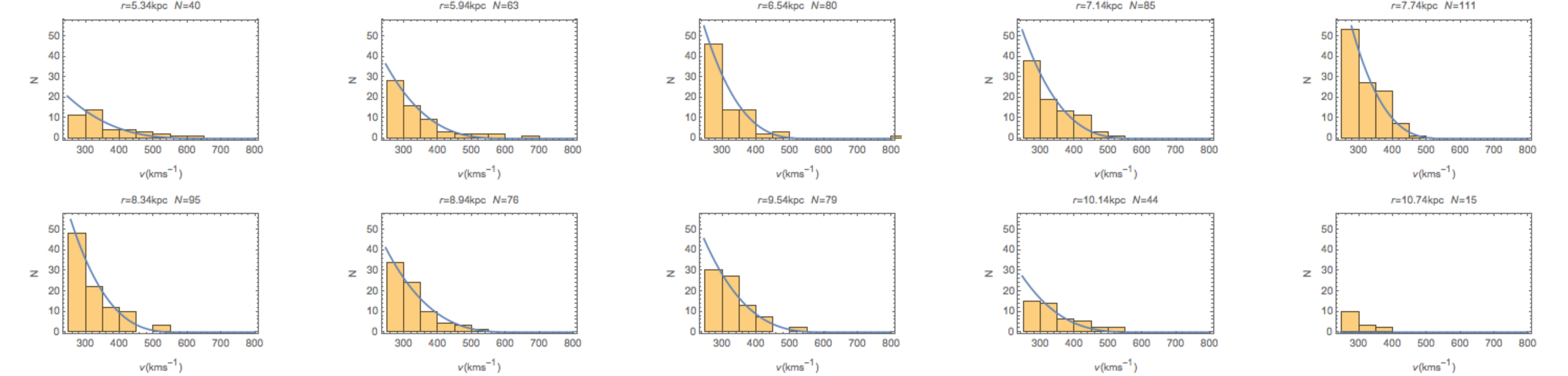}
\caption{Histograms in different $r$-bins for one realisation of $r$ and $v$ of the sample, compared with $f(v|\ve,k)$ (blue line), where $\ve$ is derived for that particular bin and $k=3$.}\label{fig:hists}
\end{figure*}

In Fig.~\ref{fig:hists} we show the histograms in different $r$-bins for one realisation of $r$ and $v$ of the sample in order to check that a power law is a reasonable description of the velocity distribution. We overplot $f(v|\ve,k)$, where $\ve$ is derived for those particular bins, and $k=3$ (the mean between $k=2.3$ and $k=3.7$, used for the marginalisation to obtain the $\ve$ probability distribution function).

\begin{figure}
\includegraphics[width=\columnwidth]{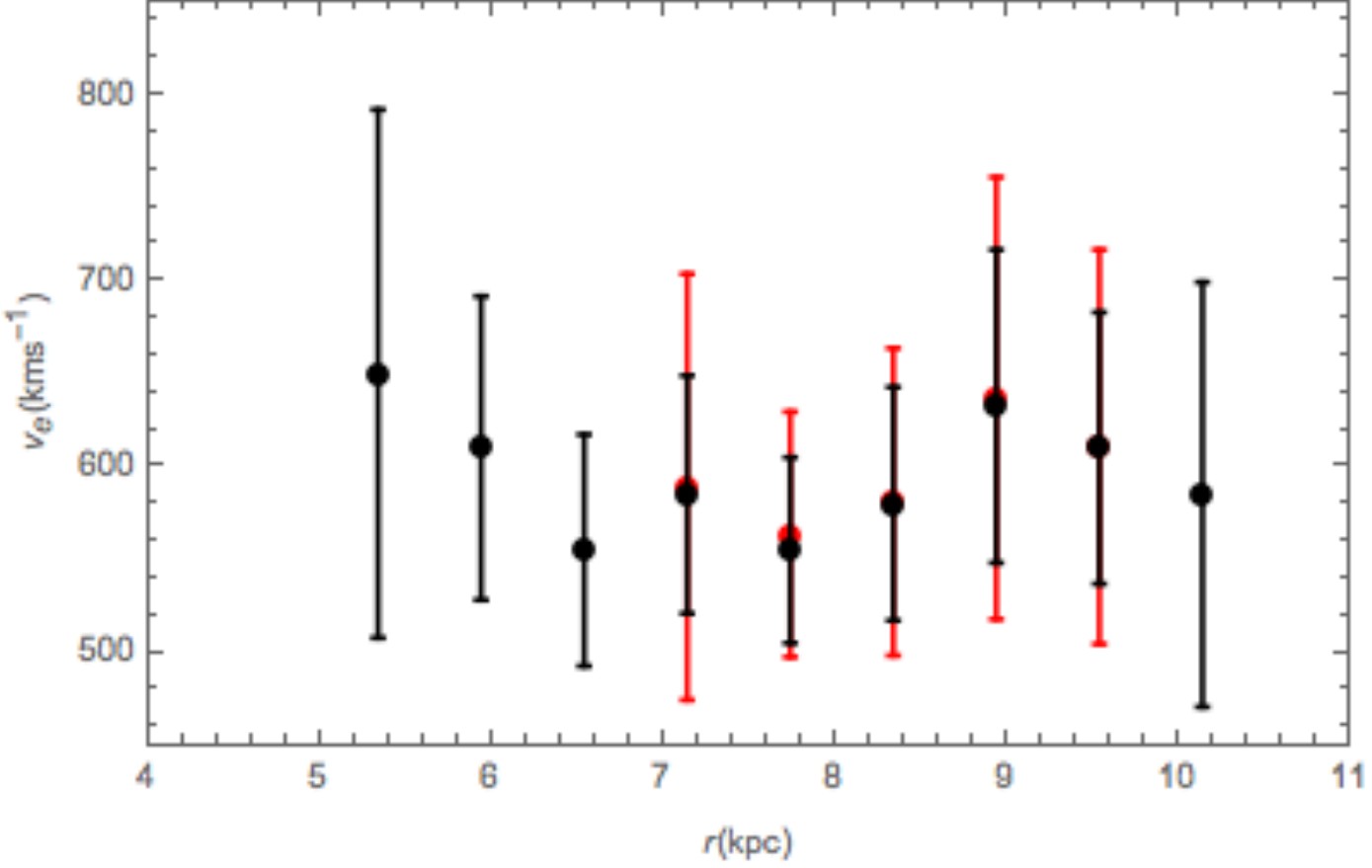}
\caption{Escape speed at different radii $r$ in the Milky Way obtained with the \texttt{StarHorse} code (black points), and McMillan distances (red points) and their error bars obtained using the procedure explained in Section~\ref{sec:Methods}.
}\label{fig:Rbins}
\end{figure}

In the left panel of Fig.~\ref{fig:Rbins}, we show the results of the analysis in different bins (black dots with error bars). Only the bins that contain at least ten stars in each resampling were considered. The escape velocity curve decreases with $r,$ as expected in any reasonable Galactic potential (see below). The value of the escape velocity at the Sun is 
\begin{equation}
\ve(r_\odot)=580 \pm 63~\kms. 
\end{equation}
The error bar increases for the farthest bins, especially the outermost bins that are less densely populated.

We then compared the results obtained with the \texttt{StarHorse} code with the distance estimates by \cite{McMillan2018}. We used the same counter-rotating stars as selected in Section~\ref{sec:data}, but with the velocities $v$ and radii $r$, computed with the new distances. The results in different bins are also shown in Fig.~\ref{fig:Rbins} with red dots and associated error bars. The results for all bins are very similar to the results obtained with the \texttt{StarHorse} distances. The escape velocity at the Sun is estimated in this case to be $\ve(r_\odot)=593\pm 76~\kms$. However, the farthest bins do not contain enough stars to perform the analysis.
With the McMillan distances it is not possible to go very far from the Sun because the distance quality degrades rapidly and the bins in our sample rapidly become empty, while the multiband photometric information used in \texttt{StarHorse} allows us to achieve accurate distances deeper inside the Galaxy.

\section{Dark matter halo mass and concentration}\label{sec:results2}

The escape velocity is in principle defined as the velocity necessary to bring a star to infinity. If $\Phi(\boldsymbol{x})$ is the potential of the Galaxy at some position $\boldsymbol{x}$, and $\Phi\rightarrow 0$ at infinity, then 
\begin{equation}
\vei(\boldsymbol{x})=\sqrt{2|\Phi(\boldsymbol{x})|}.
\end{equation}
However, it is unphysical to think that stars faster than $\ve$ derived in the previous section can reach infinity; they will instead turn around beyond some very large distance. On the basis of simulations, P14 chose this distance to be $3r_{340}$  , where $r_{340}$ is the spherical radius within which the average density of the whole Galaxy is equal to 340 times the critical density at redshift 0,
\begin{equation}
\rho_\text{c}=3H^2/8\pi G,
\end{equation}
where we take $H=73~\kms\,{\rm Mpc}^{-1}$. The distance $3r_{340}$ approximates the virial radius of the Galaxy.

\begin{figure*}
\includegraphics[width=\columnwidth]{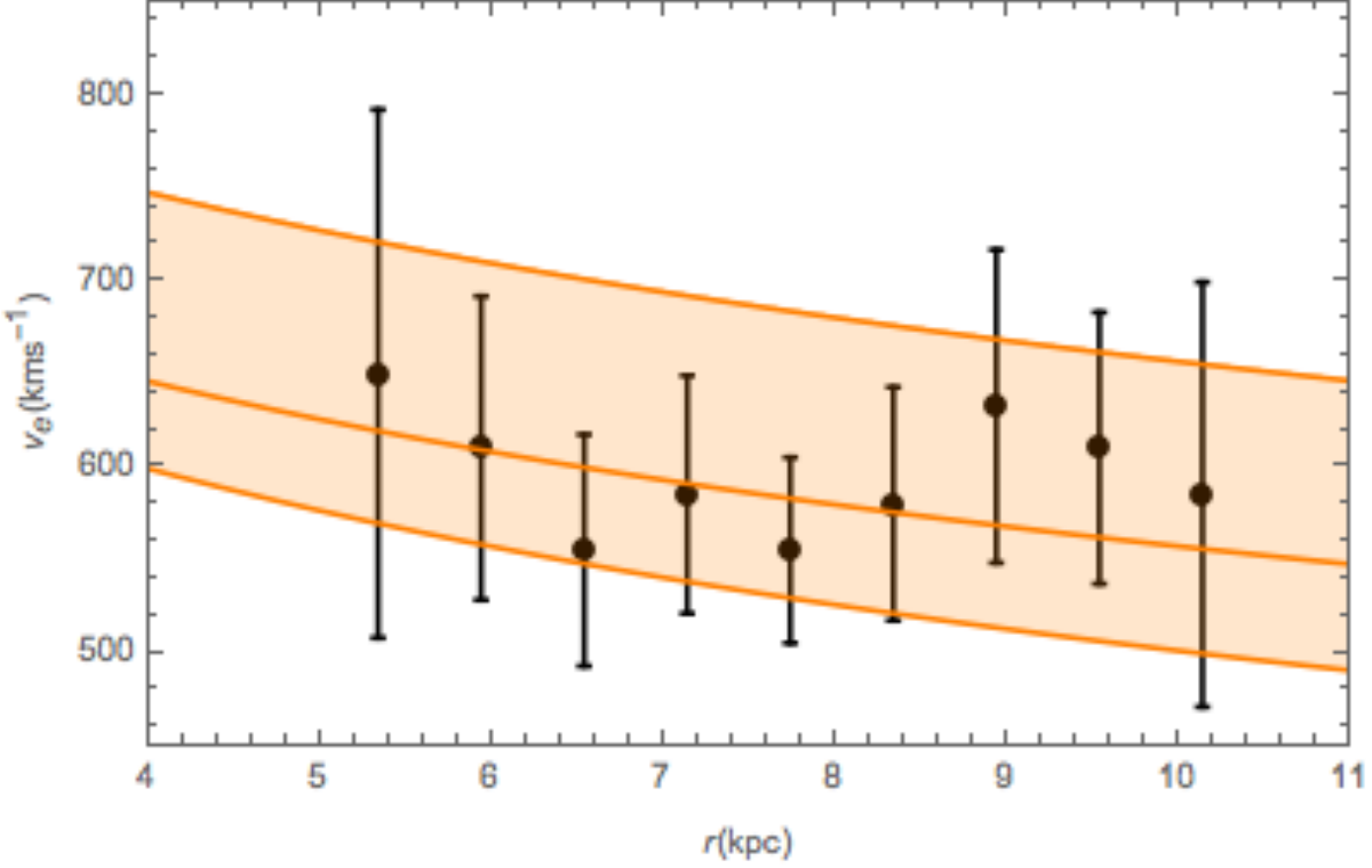}
\includegraphics[width=\columnwidth]{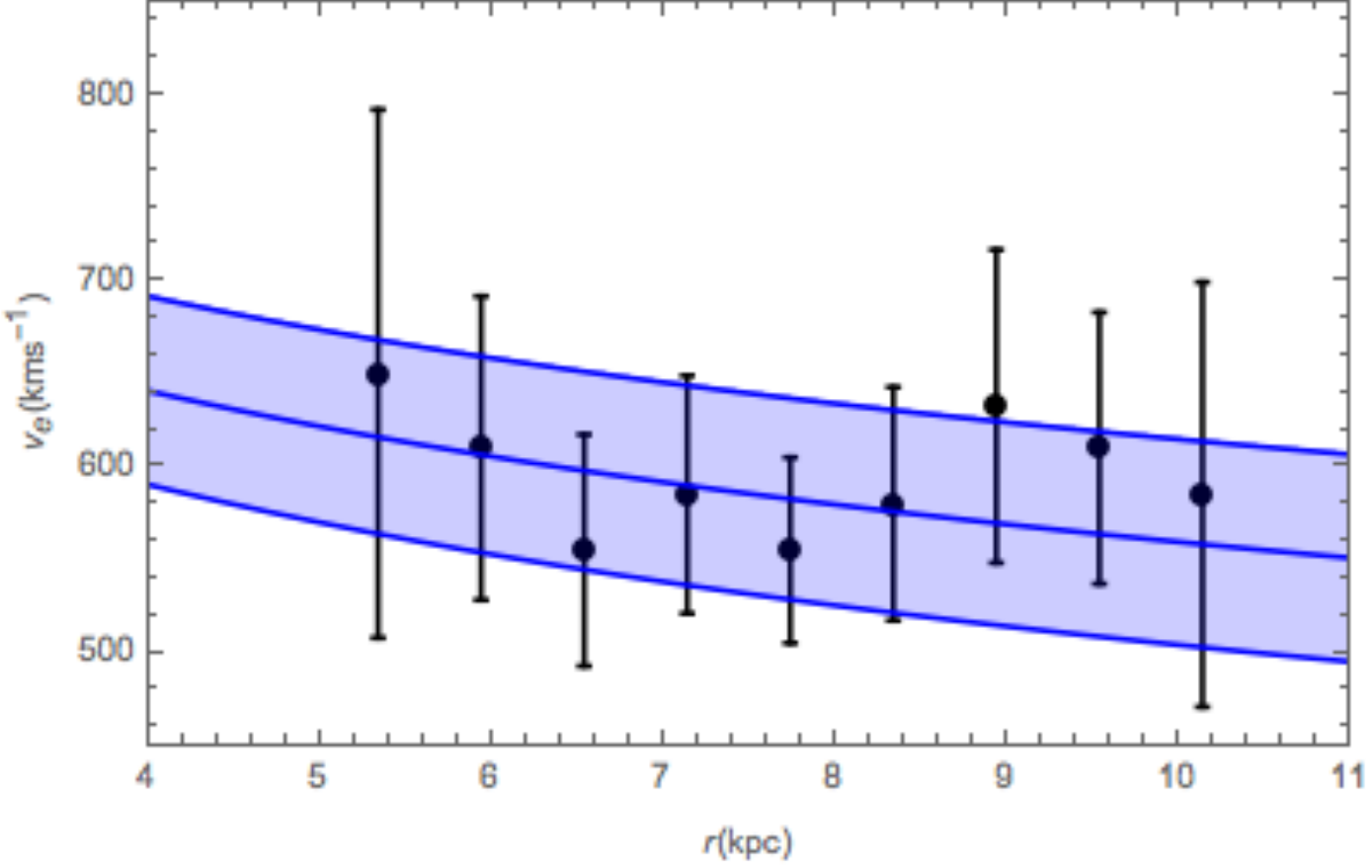}
\caption{Left panel: Fit to the escape velocity points assuming $\vetr$, and obtained using the disc and bulge potential of model III of \cite{Irrgang2013}, and a dark halo of $M_{200}=1.28\times 10^{12} M_\odot$ and $c_{200}=11.09$ (orange line and uncertainty bands). Right panel: Same as in the left panel, but for $M_{200}=1.55\times 10^{12}~M_\odot$ and $c_{200}=7.93$ (blue line and uncertainty bands) when the $\Lambda$CDM relation between $M_{200}$ and $c_{200}$ is assumed. The orange and blue bands represent the $1\sigma$ uncertainties of the models. 
}\label{fig:Rbinsfit}
\end{figure*}

To model the escape speed, we used the Milky Way potential model III of \cite{Irrgang2013}, which fits a number of observables of the Milky Way. As our estimates of the escape speed depend only on the spherical radius, we computed the Irrgang potential only within the Galactic plane\footnote{The escape speed estimated in the Galactic plane or towards the Galactic pole at a fixed $r$ can differ by a few tens of $\kms$, which is included in the error bars of our estimate of the escape speed.}, so that it depends on $r$ alone, that is, $\Phi(r)$.
The escape speed can in this case be modelled as $\vei(r)$
or as
\begin{equation}
\vetr(r)=\sqrt{2|\Phi(r)-\Phi(3r_{340})|},
\end{equation}
the latter being more meaningful in a cosmological context. We kept the Irrgang model III disc and bulge potentials fixed and varied the parameters of the NFW halo, which we chose to describe with its `virial' mass $M_{200}$ and its concentration $c_{200}$, both estimated at $r_{200}$, the radius where the average density of the dark halo reaches 200 times the critical density $\rho_\text{c}$. 
As an additional constraint, we imposed that the circular velocity at the Sun is $v_\text{c}(r_\odot)=240~\kms$, as in the assumptions that we used to transform l.o.s. velocities and proper motions to $v$. We obtain a fit using a $\chi^2$ minimisation, with the points weighted according to their error bars. The best-fit model corresponds to the orange line and orange $1\sigma$ error bands in the left panel of Fig.~\ref{fig:Rbinsfit}. The error bands are obtained using the $\chi^2$ statistics. The best-fit model has $M_{200}=0.97^{+0.47}_{-0.36} \times 10^{12} M_\odot$ and $c_{200}=12.63^{+3.41}_{-2.09}$ if the fit is made assuming the $\vei$ interpretation of the escape speed, and $M_{200}=1.28^{+0.68}_{-0.50} \times 10^{12} M_\odot$ and $c_{200}=11.09^{+2.94}_{-1.79}$ if the fit is made with the more realistic $\vetr$ (shown in the figure).

However, $N$-body simulations in $\Lambda$CDM cosmology display a relation between the halo concentrations and their mass \citep{DuttonMaccio2014},
\begin{equation}
\log_{10}(c_{200})=0.905 - 0.101\log_{10}(M_{200}/[10^{12}h^{-1}M_\odot]),
\end{equation}
where $h=H/(100~\kms~\text{Mpc}^{-1})$. We repeated the fit using this relation. In this case, the relation between $M_{200}$ and $c_{200}$ no longer guarantees that $v_\text{c}=240~\kms$ at $R_\odot$. We note that the value of the circular velocity at the solar circle is degenerate with the actual peculiar velocity of the Sun, so that a lower value of the circular velocity and a higher value of $V_\odot$ \citep[e.g.][]{Bovy} are still compatible with our assumptions. The fit is shown in the right panel of Fig.~\ref{fig:Rbinsfit} with the blue line and blue $1\sigma$ error bands. The best-fit model in this case is $M_{200}=1.16_{-0.38}^{+0.47}\times 10^{12}~M_\odot$ and $c_{200}=8.17_{-0.28}^{+0.33}$ in the $\vei$ interpretation, and $M_{200}=1.55_{-0.51}^{+0.64}\times 10^{12}~M_\odot$, $c_{200}=7.93_{-0.27}^{+0.33}$ in the more realistic $\vetr$ interpretation. In the former case, the circular speed is still well behaved in the Galaxy and corresponds to $v_\text{c}(r_\odot)=228~\kms$. Such a high mass for the Milky Way dark matter halo is expected from abundance matching. \citet{Behroozi} estimated that ${\rm 
log}(M_{200}) \simeq 12.25$ should correspond to a stellar mass of between $3 \times 10^{10} \, M_\odot$ and $5.5 \times 10^{10} \, M_\odot$.

\begin{figure}
\includegraphics[width=\columnwidth]{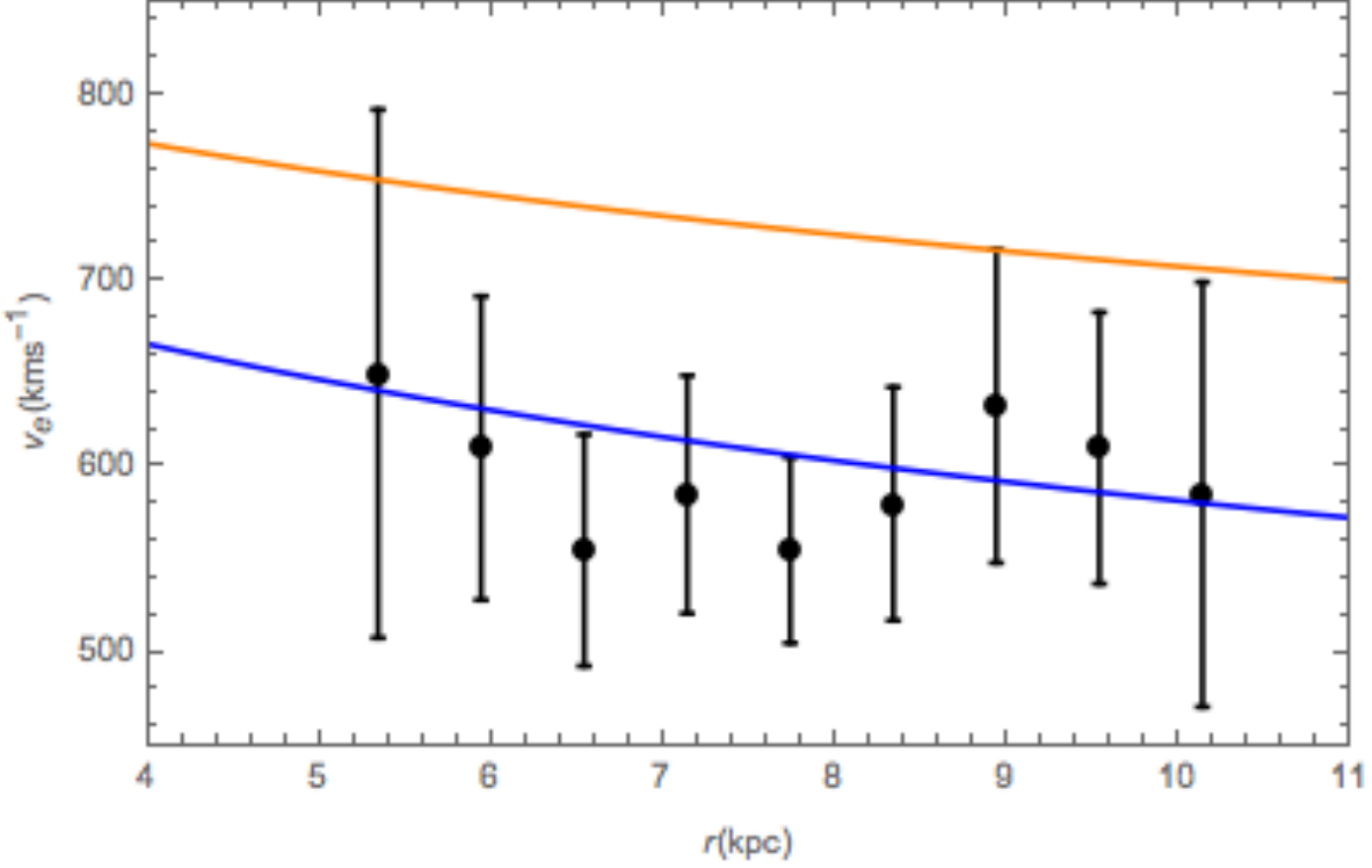}
\caption{Same as in Fig.~\ref{fig:Rbins}, but for the isothermal best fit model, with $v_0=176~\kms$ and $r_\text{c}=4.7\kpc$ (blue line), compared with the model of Englmaier \& Gerhard with $v_0=235~\kms$ and $r_\text{c}=10.7~\kpc$ (orange line). }\label{fig:RbinsIso}
\end{figure}

Finally, in Fig.~\ref{fig:RbinsIso} we show with a blue line the fit to the escape velocity points that we obtained with a quasi-isothermal potential for the dark halo of the form 
\begin{equation}
\Phi_\text{h}=\frac{v_0^2}{2}\log\left(r^2+r_\text{c}^2\right).
\end{equation}
The fit was obtained using $\vetr$ ($\Phi_\text{h}$ diverges at infinity) and imposing that $v_c(r_\odot)=240~\kms$. The best fit corresponds to $v_0=176_{-22}^{+22}~\kms$ and $r_\text{c}=4.7_{-3.8}^{+2.1}~\kpc$ (the error bars are not shown in Fig.~\ref{fig:RbinsIso}). This fit corresponds to a high halo mass $M_{200}=1.74_{-0.58}^{+0.72}\times 10^{12}~M_\odot$, as expected for a pseudo-isothermal sphere with a density decreasing slowly as $r^{-2}$ in the outer parts. We also compared our escape speed curve with the $\vetr$ values obtained from the halo model fitting the Milky Way gas dynamics by \cite{Englmaier}, with $v_0=235~\kms$ and $r_\text{c}=10.7~\kpc$, which gives a too high a value for the escape speed if no cut-off on the $r^{-2}$ is applied in the outer halo.

The analysis we performed in Section~\ref{sec:results1} can also be performed by keeping the exponent $k$ fixed. In this case, a larger $\ve$ is obtained for larger $k$ because of the correlation between these two parameters, as has been shown by \cite{LeonardTremaine1990}. To check that our estimates are consistent with such an alternative method, we derived $\ve$ for a fixed value $k=2.3$ and $k=3.7$, the extremes of the $k$-range used in our marginalisation. The $\ve$ points obtained in this way are either shifted at lower or higher values, for $k=2.3$ and $3.7$ respectively, leaving the global shape of the curve unchanged. Fitting $\vei$ and $\vetr$ to these points, we then find models that are safely included within the error bands on mass and concentration given above.

\section{Discussion and conclusions}\label{sec:conc}

We have calculated the escape speed curve of the Milky Way in a wide range of Galactocentric spherical radii using stars from the second data release of Gaia with line-of-sight velocities. The distances of these stars were computed using the \texttt{StarHorse} pipeline, which allows determining accurate and precise distances even for stars very far from the Sun thanks to its treatment of the extinction, especially in the central regions of the Galaxy. We used stars with distance estimates better than 10\% at a distance of less than $6~\kpc$ that are counter-rotating in order to have a stellar halo sample that is not contaminated by the disc. 

In the solar neighbourhood we estimated the escape speed $\ve(r_\odot)=580 \pm 63~\kms$ \citep[in $1\sigma$ agreement with][]{Williams2017}. The escape speed varies from $\sim 650~\kms$ at $\sim 5~\kpc$ to $\sim 550~\kms$ at $10.5~\kpc$. The uncertainty in the determination of $\ve(r_\odot)$ is quite large despite the relatively large stellar sample, mainly because this uncertainty takes into account the errors in position and velocity of the stars in the analysis, and treats the possibility that they might be bound or not in the Galactic potential with a bootstrap technique without excluding stars on arbitrary criteria. Moreover, our determination of the escape speed at the Sun and at other radii in the Galaxy is `local' (i.e. performed in small radial bins), meaning that the estimates do not require any velocity rescaling that assumes a Galactic potential {\it \textup{a priori}} (see P14); this introduces a bias in the result. 
Our determination of $\ve(r_\odot)$ is slightly larger than but agrees well with what P14 determined based on stars from the RAVE survey, within the error bars. Our result also agrees with the finding by \cite{Hattori2018}, who reported that the escape speed in the solar neighbourhood is expected to be of the order of $\sim 600~\kms$.

We have modelled the escape speed data points obtained using simple forms for the Galactic potential, including a fixed disc and bulge, by fitting a number of observables and a dark halo specified by the free parameters for the fit. The escape speed can be interpreted either as the velocity necessary to bring a star to infinity from a location in the Galaxy, or, more realistically, to a typical radius very far from the Galactic centre, motivated by cosmological simulations. The latter is chosen to be $3 r_{340}$ (as in P14). In the latter way, and fixing the circular velocity of the Galaxy at $240~\kms$ at the Sun, $M_{200}=1.28^{+0.68}_{-0.50} \times 10^{12} M_\odot$.

With this fit imposing the local circular velocity at $240~\kms$ , we find a slightly more concentrated halo than in $\Lambda$CDM simulations. However, when we performed the fit imposing the $\Lambda$CDM relation between halo masses and concentrations, we find a higher halo mass ($M_{200}=1.55_{-0.51}^{+0.64}\times 10^{12}~M_\odot$), a concentration $c_{200}=7.93^{+0.33}_{-0.27}$ \citep[e.g.][]{Kruijssen2018}, and still a reasonable circular velocity curve. 
We also fit models consisting of the same disc and bulge potentials, but including cored pseudo-isothermal dark halos. In this case, the only interpretation for the escape velocity is the velocity to bring a star at $3r_{340}$, and we find $M_{200}=1.74\times 10^{12}~M_\odot$, with a large core radius $r_\text{c}=4.7~\kpc$. In the future, we aim to determine whether such a measurement is compatible with alternative models \citep[e.g.][]{Famaey2007}

The high estimated mass of the Milky Way can be compared with recent estimates of the total mass of the Local Group based on the timing argument \citep{Penarrubia}, which sets the total mass to $\sim 2.64^{+0.42}_{-0.38}\times 10^{12}~M_\odot$. The heavy Milky Way implied by our work seems in accordance with recent estimates of the mass of M31, yielding a lower mass than previously thought \citep[$0.8\pm 0.1\times 10^{12}~ M_\odot$,][]{Kafle2018}, which leaves room for an increased Milky Way mass.

A caveat of this work is that the distance determination for stars using Gaia stellar parameters is still an ongoing process, and the distances and velocities of the stars could undergo modifications, especially at large distances \citep[e.g. because of the magnitude-dependent offset in the Gaia $G$ magnitudes, see][]{Casagrande2018}. A dynamical caveat is that the analysis with which we estimated the escape velocities is based on distribution functions and assumes phase-mixing for the kinematic tail of the stellar halo, which is not guaranteed. However, the choice of the velocity distribution is motivated by simulations, and the comparison of the derived distribution functions with the histograms of the velocity distribution of the halo stars that we performed shows that this description is reasonable. Finally, the models of the Milky Way potential that we presented here are rather simple and do not take into account a number of issues, such as the fact that the dark halo should self-consistently adapt its shape to the bulge and disc potential \citep[e.g.][]{ColeBinney2017} and that the fit could be performed leaving the disc and bulge parameters free. Such detailed modelling is beyond the scope of this paper, however. Finally, spectroscopic follow-up of the stars we used in this analysis might be used to study whether their properties are compatible with the chemical properties of the stellar halo \citep[e.g.][]{HawinsWyse2018}.

In conclusion, the uncertainties in the measurements of the escape velocity (and of the corresponding dark halo masses) are quite large, mostly because we were realistic and considered the uncertainties on both distance and velocity of the stars, and because we lack information on whether high-velocity stars are bound or unbound in the Galactic potential. Nevertheless, our estimate of the escape speed curve for a wide range of Galactocentric radii in the Milky Way conclusively implies a massive Galaxy $M_{200}>10^{12} M_\odot$, especially when considering the mass-concentration relation of $\Lambda$CDM, in agreement with other recent estimates using independent methods, both dynamical and using chemical tagging \citep[e.g.][]{HawinsWyse2018}. It also agrees with typical abundance-matching expectations. 

\begin{acknowledgements}
We thank Paul McMillan for useful discussions. RFGW thanks her sister, Katherine Barber, for support, and the Leverhulme Trust for a Visiting Professorship at the University of Edinburgh, held while this work was being completed.
We thank the E-Science and Supercomputing Group at Leibniz
Institute for Astrophysics Potsdam (AIP) for their support with running the \texttt{StarHorse} code on AIP cluster resources.
This work has made use of data from the European Space Agency (ESA)
mission {\it Gaia} (\url{https://www.cosmos.esa.int/gaia}), processed by
the {\it Gaia} Data Processing and Analysis Consortium (DPAC,
\url{https://www.cosmos.esa.int/web/gaia/dpac/consortium}). Funding
for the DPAC has been provided by national institutions, in particular
the institutions participating in the {\it Gaia} Multilateral Agreement.
\end{acknowledgements}

\bibliographystyle{aa}
\bibliography{vescbib}

\end{document}